\documentclass[aps,prb,preprint,amsbsy,amssymb]{revtex4}
\begin{document}
\draft
\title{ Electromagnetic properties of a Fermi liquid superconductor with singular Landau parameter $F_{1s}\rightarrow-1$}
\author{Tai-Kai Ng}
\address{
Department of Physics,
Hong Kong University of Science and Technology,
Clear Water Bay Road,
Kowloon, Hong Kong
}
\date{ \today }
\begin{abstract}
  The electromagnetic properties of a Fermi liquid superconductor with singular Landau parameter $F_{1s}\rightarrow-1$,
 corresponding to a state with marginal spin-charge separation (MSCS) is analyzed in this paper. We show that the
 MSCS state describes a strongly phase-disordered superconductor which is a diamagnetic metal with Drude-like
 optical conductivity. The phase diagram and electromagnetic properties of the MSCS state is found to be in
 qualitative agreement with what is observed in the pseudo-gap phase of (underdoped) High-$T_c$ cuprates. We
 predicted that vortices with {\em un-quantized} magnetic flux will be observed in the pseudo-gap phase.

\end{abstract}


\maketitle

    The understanding of the normal state of underdoped cuprates (pseudo-gap phase) remains one of the major theoretical
 challenge to the condensed matter physics community nowadays because the phase seems to violate a number of
 properties required by Fermi liquid theory\cite{r1,r3,r4}. On the other hand, the superconducting phase, which
 is just below the pseudo-gap phase in the $T-\delta (\delta=$ hole concentration) phase diagram, seems to obey
 Fermi-liquid (superconductor) phenomenology pretty well\cite{r1,r3,r4,t1,ng1}. The proximity of the two
 phases suggests that it may be possible to describe the pseudo-gap phase starting from an effective Fermi-liquid
 theory that describes the superconducting state of the cuprates. This scenario, if correct, implies that we may
 be able to develop an unified phenomenological framework to understand both the pseudo-gap and superconducting
 phases in underdoped cuprates.

   Recently, based on an analysis of the slave-boson mean-field theory (SBMFT) of $t-J$ model, we proposed that the
 pseudo-gap state can be described as a Fermi liquid superconductor with singular Landau parameter $F_{1s}\rightarrow-1$,
 corresponding to a state with spin-charge separation in the long wave-length, low frequency limit\cite{chan-ng}. We
 shall call this state a marginally-spin-charge separated state (MSCS) in the following. Superfluidity disappears
 in the MSCS state because quasi-particles do not carry current in this limit\cite{chan-ng}. In this paper we
 shall examine the electromagnetic properties of the MSCS state in more details by considering a general Landau
 Fermi liquid superconductor with Fermi liquid parameter $F_{1s}\rightarrow-1$, but with other Fermi-liquid
 parameters remaining regular. We shall show that the MSCS state describes a strongly phase-disordered
 superconductor which is a diamagnetic metal with Drude-like optical conductivity. A prediction on the appearance
 of vortices with un-quantized magnetic flux is made.

     In the MSCS scenario, the pseudo-gap phase is a Fermi-liquid superconductor with a parent BCS superconductor
 mean-field transition temperature $T_M\sim T^*$(pseudo-gap temperature). The current carried by quasi-particles
 are renormalized by the Landau parameter $1+F_{1s}$\cite{r3,r4,t1,ng1} and vanishes in limit
 $F_{1s}\rightarrow-1$\cite{chan-ng}. We shall consider in the following a $d$-wave superconductor with Landau
 parameter $1+F_{1s}(T)=z>0$ at temperature $T=0$ and reduced gradually to $1+F_{1s}(T)\rightarrow0$ at
 $T=T_c<T_M$. We note that translational and rotational symmetries are strongly broken by the Fermi surface
 geometry in cuprates, and the Landau interactions $F_{\vec{k}\vec{k}'}^{s(a)}$ depend in general not only on the
 relative angle between $\vec{k}$ and $\vec{k}'$, but also on the directions of the $\vec{k}$ and $\vec{k}'$
 vectors themselves\cite{pines}. This is not taken into account in our simply parametrization of Landau parameter
 since the aim of our paper is not to establish a complete Fermi liquid theory for cuprates, but to examine the
 general consequences of $F_{1s}\rightarrow-1$. We note also that there exists no general relation between the
 effective mass $m^*/m$ and $F_{1s}$ for a system with broken translational symmetry\cite{pines}. For an ordinary
 (non-superconducting) Fermi liquid which breaks translational symmetry, the point $F_{1s}=-1$ is a
 critical point\cite{pines} separating the Fermi liquid state ($F_{1s}>-1$) and a state with spontaneous current
 ($F_{1s}<-1$) and magnetic flux. However, for a Fermi liquid superconductor, this instability is suppressed by
 the opening of gap on the Fermi surface.

    When $F_{1s}\rightarrow-1$, we have to include finite wave-vector and frequency corrections to the Landau
 interaction and replaces $F_{1s}$ by a Landau function $F_{1s}(q,\omega)$. Assuming that $F_{1s}(q,\omega)$
 is regular at $q,\omega\rightarrow0$, we expect $F_{1s}(q,\omega)\sim-1+\chi_d(T)q^2+K(\omega)$ in the limit
 $q<<k_F,\hbar\omega<<T_M$, where $\chi_d(T)>0$ (stability requirement for Fermi liquids) and $K(\omega)$ are
 phenomenological parameter and function to be determined. We shall see later that $\chi_d(T)$ measures
 the magnetic susceptibility of the system and $\chi_d>0$ implies that the system is diamagnetic. In particular,
 at $T\rightarrow T_c^+$ we expect
 \begin{equation}
 \label{chi}
 \chi_d(T)\sim z{r_M^2\over(T/T_c-1)^\nu},
 \end{equation}
 where $\nu>0$ and $r_M$ is a characteristic length scale above which spin-charge separation occurs.
 Physically the (diamagnetic) susceptibility should diverge as the system approach the superconducting state. We
 shall assume also that $K(\omega)$ has the form
 \begin{mathletters}
 \label{kk}
 \begin{equation}
 \label{k1}
   Re\{K(\omega)\}=z({\omega^2\over\omega^2+\Gamma(T)^2}),
 \end{equation}
 so that $1+F_{1s}\rightarrow0$ at $\omega\rightarrow0$, and $\rightarrow z$ at $\omega>>\Gamma(T)$, i.e. the quasi-particles
 "recover" their charges at $\omega>>\Gamma(T)$. We shall see later that the recovery of charges at large $\omega$
 is "required" by the $f$-sum rule. Notice that
 \begin{equation}
 \label{k2}
   Im\{K(\omega)\}=-{z\Gamma(T)\omega\over\omega^2+\Gamma(T)^2}\neq0  ,
 \end{equation}
 \end{mathletters}
 by the Kramers-Kronig relation. $T_M, T_c, z, \nu, r_M$ and $\Gamma(T)$ are phenomenological parameters to be
 determined from experimental datas on the underdoped cuprates. We shall assume $T_M>>T_c\sim\delta$ in the
 following analysis.

   Our simple phenomenology has some non-trivial consequences. To see that we examine the current-current
 response function in the MSCS state. The (transverse) current-current response function $\chi_T(q,\omega;T)$
 at temperature $T$ for a Fermi-liquid superconductor is given by\cite{chan-ng,leggett}
 \[
 \chi_T(q,\omega;T)={\chi_{0t}(q,\omega;T)\over1-\left({F_{1s}(q,\omega)\over1+F_{1s}(q,\omega)}\right)
  {\chi_{0t}(q,\omega;T)\over\chi_{0t}(0,0;0)}},
 \]
 where $\chi_{0t}(q,\omega;T)$ is the transverse current-current response function of the corresponding BCS
 superconductor in the absence of Landau interactions. Using our phenomenological form of
 $F_{1s}(q,\omega)$, we obtain at small $q$ and $(\omega,k_BT)<<k_BT_M$,
 \begin{equation}
 \label{cc1}
 \chi_T(q,\omega;T>T_c)\rightarrow\left(\chi_d(T)q^2+K(\omega)\right)\chi_{0t}(0,0;0)+O({T\over T_M},
 {\hbar\omega\over k_BT_M}),
 \end{equation}
  where the $O({T\over T_M},{\hbar\omega\over k_BT_M})$ terms are contributions from quasi-particle excitations.
  At $\omega,T<<T_M$, the quasi-particle contributions can be neglected and we find that the MSCS state is a
 diamagnetic metal with magnetic susceptibility $-\chi_d(T)$ and AC conductivity
 \begin{equation}
 \label{ac}
 \sigma_d(\omega)\sim{K(\omega)\chi_{0t}(0,0;0)\over(-i\omega)}={z\chi_{0t}(0,0;0)\over\Gamma(T)-i\omega},
 \end{equation}
 corresponding to a Drude metal with carrier density $\sim z$ and inverse life time $\Gamma(T)$. Notice that
 there is no "quasi-particle" contribution to the AC conductivity in our approximation.

   This rather surprising result can be better understood by examining the integral
 \[
  I=2\int_0^{\infty} {d\omega\over\pi}Re[\sigma(\omega)]d\omega,  \]
  which, according to the $f$-sum-rule\cite{pines}, is a number depending on the total charge density only and is
  independent of whether the system is in the superconducting or normal state. In the superconducting state at $T=0$,
  $I=I_s+I_Q$ where $I_s=(1+F_{1s})\chi_{0t}(0,0;0)=z\chi_{0t}(0,0;0)$ is the superfluid contribution to
  $I$\cite{leggett,tinkham}. $I_Q$ is the contribution from quasi-particle excitations. Assuming $T<<T_M,
  \hbar\Gamma(T)<<k_BT_M$ and the vanishing of superfluidity is driven by $F_{1s}\rightarrow-1$ with the
  quasi-particle contributions $I_Q$ remaining more or less unaltered, we obtain
  approximately from Eq.\ (\ref{ac})
  \begin{equation}
  \label{sr1}
   {I_s\over\chi_{0t}(0,0;0)}\sim2\int_0^{\infty}{d\omega\over\pi}{Re[\sigma_d(\omega)]\over\chi_{0t}(0,0;0)}=
   2\int_0^{\infty}{d\omega\over\pi}{Im[K(\omega)]\over0-\omega}=(Re[K(\infty)]-Re[K(0))],
  \end{equation}
   which says that the missing spectral weight when the system goes from the superconducting state to the
   pseudo-gap state should "re-appear" in the difference between $K(\infty)$ and $K(0)$ in order to satisfy the
   $f$-sum rule. In the MSCS state where $1+F_{1s}(0,0)=0$, we must have $1+F_{1s}(0,\infty)\rightarrow z$.
   This is precisely what we have chosen in our phenomenological form of $K(\omega)$. A corresponding
   contribution to AC conductivity appears indicating that additional charge degree of freedom not accountable by
   quasi-particle excitations has to appear in the MSCS state because of charge conservation, i.e. spin-charge
   separation. The precise form of $K(\omega)$ is determined by the microscopic dynamics of this "anomalous"
   charge field which cannot be determined by our simple Fermi-liquid phenomenology\cite{chan-ng}. Nevertheless
   analyticity suggests that a Drude-like conductivity is likely to occur, as indicated by our simply
   parametrization of $K(\omega)$.

   To study in more details the properties of the system we derive the Ginzburg-Landau (GL) free
  energy for the MSCS state. Following Ref.\cite{ngtse}, we consider a Fermi-liquid superconductor with Landau
  parameters $F_{0s}$ and $F_{1s}$ only. Writing the superconductor order parameter $\psi$ in terms of amplitude
  and phase variables $\psi=\sqrt{\rho}e^{i\theta}$, it was found that the GL free energy at $T<<T_M$ is\cite{ngtse}
  \begin{equation}
  \label{gleff}
  F_{GL}\sim\int d^dx \left({\hbar^2\over2m^*}(\nabla\sqrt{\rho})^2-\alpha(T)\rho+{\bar{\beta}\over2}\rho^2
    +{\hbar^2\over2m^*}(1+F_{1s})\rho(\nabla\theta-{2\pi\over\Phi_0}\vec{A})^2+{\vec{B}^2\over8\pi}
    -{\vec{B}.\vec{H}\over4\pi}\right).
  \end{equation}
  where $\bar{\beta}=(1+F_{0s})\beta$. $\alpha(T)$ and $\beta$ are the usual parameters parameterizing the GL
  free energy of the BCS superconductor without Landau interactions. $\vec{A}$ is the vector potential, $\vec{B}=\nabla\times
  \vec{A}$ is the total magnetic field and $\Phi_0=hc/2e$ is the fluxoid quantum. $\vec{H}$ is the external
  magnetic field. $F_{0s}$ and $F_{1s}$ appear in the renormalized parameter $\bar{\beta}=(1+F_{0s})\beta$ and
  superfluid density $\rho_s=(1+F_{1s})\rho$, in agreement with Fermi liquid theory\cite{leggett,ngtse}. The
  (mean-field) MSCS state is obtained by replacing $1+F_{1s}\sim z$ at $T=0$ by $1+F_{1s}\sim-\chi_d\nabla^2$ at
  $T>T_c$ with correspondingly,
  \[
  (1+F_{1s})\rho(\nabla\theta-{2\pi\over\Phi_0}\vec{A})^2\rightarrow
  \chi_d\rho\left((\nabla^2\theta)^2+4\pi^2(\vec{\kappa}-{\vec{B}\over\Phi_0})^2\right),
  \]
   where $\vec{\kappa}={1\over2\pi}\nabla\times\nabla\theta$ is the vorticity. Notice the vanishing of superfluid
   rigidity and disappearance of Meissner effect in the MSCS state. We note also that the resistive response to
   electric field will be recovered if we consider the time-dependent GL action\cite{ngtse} with finite frequency
   responses included in $F_{1s}$.

     At $T<<T_M$, amplitude fluctuations of the superconducting order parameter is unimportant
     and we can replace the GL free energy by an effective phase free energy
   \begin{equation}
   \label{phase}
     F_{\theta}=\int d^dx\left({\hbar^2\bar{\rho}\chi_d\over2m^*}[(\nabla^2\theta)^2+4\pi^2
   (\vec{\kappa}-{\vec{B}\over\Phi_0})^2]-{\vec{B}^2\over8\pi}-{\vec{B}.\vec{H}\over4\pi}+
   {\epsilon_v\over2\pi\xi^2}[\vec{\kappa}]^2\right),
   \end{equation}
    where $\bar{\rho}=\alpha(T)/\bar{\beta}$, $\epsilon_v$ is the vortex core energy and $\xi$ is the coherence
    length. Notice that although Meissner effect disappears, a residual interaction between vortices and magnetic
    field still exists which is proportional to $\chi_d(T)$. Minimizing the free energy with respect to $\vec{B}$,
    we find that a vortex still bind a total magnetic flux
    \begin{mathletters}
    \label{vortex}
    \begin{equation}
    \label{v1}
     \Phi_v\sim{(\chi_d(T)\lambda^{-2})\over1+(\chi_d(T)\lambda^{-2})}\Phi_0,
    \end{equation}
     where $\lambda$ is the London penetration depth of the parent BCS superconductor in the absence of Landau
     interaction. The total vortex energy in the presence of external field $\vec{H}$ is
     \begin{equation}
     \label{v2}
     \bar{\epsilon}_{vs}(H)=\epsilon_v\left(1+({a'\chi_d(T)\xi^{-2}\over1+(\chi_d(T)\lambda^{-2})})
     (1-{2Hs\over h_{c2}})\right),
     \end{equation}
     \end{mathletters}
     where $s=\pm1$ is the sign of the vorticity and $h_{c2}\sim\Phi_0/2\pi\xi^2$ is the upper critical field
    for the parent BCS superconductor. $a'$ is an numerical factor of order $O(1)$. The absence of long-range
    interaction between vortices suggest that the MSCS state is a strongly disordered vortex liquid,
    with vortex density $n_{s}(T)\sim\xi^{-2}exp\left(-{\bar{\epsilon}_{vs}(H)\over k_BT}\right)$\cite{r3,bkt,t4}.
    Notice that the above estimates become inaccurate at the critical point $T\sim T_c$ when
    $\chi_d(T)n_s(T)\geq1$(see Eq.\ref{chi})).

      The phase diagram of the system can be determined if we assume that $1+F_{1s}=a(1-T/T_c)$ at $T<T_c$, where
    $a\sim z$ and the free energy is given by the usual $x-y$ model phase action\cite{bkt,t4}
    \begin{equation}
    \label{f2}
      F(T<T_c)\sim a(1-{T\over T_c}){\bar{\rho}\hbar^2\over2m^*}\int d^dx(\nabla\theta-{2\pi\over\Phi_0}\vec{A})^2.
    \end{equation}

      Assuming that the system is quasi-two dimensional we expect that superconductivity will exist at
    $T<T_{KT}\sim \gamma T_c$, where $T_{KT}$ is the Berezinskii-Kosterlitz-Thouless (BKT) transition\cite{bkt,t4}
    temperature and $\gamma\sim{(a\bar{\rho}\hbar^2/4m^*T_c)\over(1+(a\bar{\rho}\hbar^2/4m^*T_c)}<1$. The system
    goes through a BKT transition at $T_{KT}$ into the vortex liquid state which is a normal metal with strong
    superconductivity fluctuations and para-conductivity behavior\cite{p1}. This region can be identified with the
    region with strong Nernst signal in the underdoped cuprates\cite{o1,o2}. The system cross-overs into the MSCS
    state at $T\geq T_c$ where phase rigidity is lost completely. Notice that
    the mean-field superconducting-to-MSCS transition is superseded and smeared out by the BKT transition into a
    crossover. Vortex properties in the MSCS state is anomalous. The magnetic flux trapped in the vortex core
    is {\em un-quantized} and decreases gradually when the system crossover to the MSCS state (Eq.\ (\ref{v1})).

     We next discuss briefly transport properties in the MSCS state. we find that the total electrical
    conductivity $\sigma$ of the system is given by the Ioffe-Larkin composition rule\cite{il} $\sigma^{-1}=
    \sigma_d^{-1}+\sigma_f^{-1}$, where $\sigma_d$ is the anomalous (Drude) conductivity and $\sigma_f=\sigma_v+
    \sigma_Q$ is the conductivity from the "Fermi liquid" component. $\sigma_Q$ is the contribution from excited
    quasi-particles whereas $\sigma_v$ is the vortex (superfluid) contribution. Vortex contributions to
    transport can be determined if we assume that vortex dynamics is the same as vortex dynamics in usual
    superconductors\cite{tinkham,p1} except that the magnetic flux trapped inside the vortex core is $\Phi_v$.
    Notice that this assumption implies that transport in the vortex liquid state at $T<T_c$ crossover smoothly
    to the MSCS state at $T>T_c$. With this assumption the two equations governing vortex dynamics are
   \begin{eqnarray}
   \label{vdy}
   \eta\vec{v}_s+\eta_Ms\hat{z}\times\vec{v}_s & = & \vec{j}\times{s\Phi_v\hat{z}\over c}-
   ({\partial f_s\over\partial T})\nabla T,    \\  \nonumber
   \vec{E} & = & {\Phi_0\hat{z}\over c}\times(n_+\vec{v}_+-n_-\vec{v}_-)
   \end{eqnarray}
   where $\vec{v}_s$ is the velocity of a vortex with vorticity $s$, $f_s$ is the free energy associated with a vortex.
   $\eta$ is a viscous damping coefficient and $\eta_M$ is the Magnus parameter. $\vec{j}$ is the electric current and
   $\vec{E}$ is the electric field induced by vortices motion. The magnetic field is along $+\hat{z}$ direction.
   Notice that the second equation, which is the Josephson relation, represents phase slip processes\cite{tinkham}
   when a vortex pass through a line in space and is not affected by the actual amount of magnetic flux trapped
   in vortex core.

   Solving the equations it is easy to show that vortices contribution to transports are important only when
   the coupling between vortices and external magnetic field $\sim\chi_d(T)$ is large, and decreases rapidly when
   $T>>T_c$. Therefore for $T_M>>T_c$ we expect that $\sigma_f$ is dominated by $\sigma_v$ at low
   temperature regime $T\sim T_c$, and crossovers to a regime dominated by quasi-particle contributions
   $\sigma_Q$ when $T>>T_c$. In the following we shall examine the low temperature regime where
   $\sigma_Q$ is negligible.

    Assuming a Bardeen-Stephen type model\cite{bs} for $\eta$ with $\eta>>\eta_M$, we obtain for the ($\vec{B}=0$) total
    resistivity and Nernst signal ($e_N=|\vec{E}|/|\nabla T|$)
   \begin{eqnarray}
   \label{resistivity}
   \sigma^{-1} & = & \sigma_d^{-1}(1+b'(\xi^2N_+(T)){\Phi_v\over\Phi_0}),  \\ \nonumber
    e_N & =  & {2e\over\hbar}({\partial f_s\over\partial T})(\xi^2N_-(T))\sigma_d^{-1}+e_{Nd}.
   \end{eqnarray}
   where $b'$ is a numerical factor of order $O(1)$ and $N_{\pm}(T)=n_+(T)\pm n_-(T)$. $e_{Nd}$ is the contribution to
   Nernst signal from the anomalous charge dynamics which we cannot determine without additional assumptions. For
   the same reason we cannot determine with certainty $\sigma_{xy}$. Notice that vortex contribution to conductivity
   and Nernst signal decreases when temperature increases. In particular, we find that for $\chi_d(T)\xi^{-2}<<1$,
   \[
   e_N\sim{4\over c}({2e\over\hbar})^2({\partial f_s\over\partial
   T})\sigma_d^{-1}e^{-\epsilon_v\over k_BT}{\chi_d(T)\over k_BT}H+e_{Nd},  \]
   which suggests that the dominant temperature dependence of $e_N/H$ is proportional to $\chi_d(T)/k_BT$ at the
   temperature regime $T\sim$ {\em a few} $k_BT_c$ where the Nernst signal decreases rapidly with temperature, if
   we assume that $e_{Nd}$ is small and negligible and $\epsilon_v$ is of order of $k_BT_c$\cite{r3}, in qualitative
   agreement with what is observed experimentally\cite{o1,o2} in underdoped cuprates.

     Summarizing, we propose in this paper that the pseudo-gap state in underdoped cuprates is a state with marginal
   spin-charge separation (MSCS). The general properties of the MSCS state, including the appearance of Drude
   conductivity, the natural appearance of low temperature Nernst regime and high temperature regime where
   paraconductivity behavior disappears, seem to agree qualitatively with what is observed experimentally although
   detailed comparison/fitting with experimental data is not made in this paper. An important prediction of the
   theory is that vortices with {\em un-quantized} magnetic flux $\Phi_v<\Phi_0$ which decreases gradually with
   increasing temperature exist in the MSCS state.  This prediction can be tested in rings of multiple grain
   boundary Josephson junction in the pseudo-gap state\cite{tsuei} and provides an unambiguous test to the theory.

      A microscopic realization of the MSCS state is the non-Bose-condensed state in the $U(1)$
   gauge theory for the $t-J$ model\cite{ng1,chan-ng}. In this case, the anomalous charge dynamics is described
   by a charge $q=1$ complex scalar field. We note that other possibilities exist. For example, the charge
   dynamics in the spin-charge separated state will be very different in the $SU(2)$ theory\cite{t3} for the $t-J$
   model. The phenomenology presented here provided a plausible general framework to understand all these
   theories which predict spin-charge separation. More detailed investigations of properties of the MSCS state and
   comparison with experiments will be presented in future papers.

  \acknowledgements
  This work is supported by HKUGC through grants 602705 and CA05/06.SC04.

\end{document}